# Comment to "Multipolar expansion of the electrostatic interaction between charged colloids at interfaces"

*Tai-Kai Ng and Yi Zhou*

*Department of Physics, Hong Kong University of Science & Technology, Clear Water Bay Road, Hong Kong*

In a recent paper, Domínguez *et. al.*[1] claimed that the electrostatic interaction between charged colloids trapped at an interface formed by a dielectric and a screening phase is always isotropic to order $d^{-3}$, where *d* is the distance between the colloids. Base on this result, they claimed that in-plane dipolar attraction of order $d^{-3}$ between colloids cannot exist, in contrast to previous proposals [2].

The argument of Domínguez *et. al.* was based on a vital assumption, that on a flat surface, the electrostatic potential $\phi(r, z = 0, \vartheta)$ can be expanded in a series of Bessel functions $J_m(kr)$ in cylindrical coordinate, i.e.,

$$\phi(r, z = 0, \vartheta) = \sum_m \int dk \phi^m(k) J_m(kr) e^{im\vartheta} \tag{1}$$

with the coefficients $\phi^m(k)$ being a regular function at small *k*. In the presence of colloids, it was assumed that the same expansion remains valid at distance far away from the colloids. The long distance behavior of $\phi(r, z, \vartheta)$ at $z \to 0$, $r \to \infty$ can be studied by the small-k behaviors of $\phi^m(k)$, and provided that $\phi^m(k)$ is regular at small *k*, the problem reduces to analysis of integrals of form

$$\phi_j^m = \sum_m \int dk k^j J_m(kr), \tag{2}$$

which can be performed analytically. By analyzing the coefficients $\phi_j^m$, it can be shown that the electrostatic interaction between charged colloids trapped at an interface formed by a dielectric and a screening phase is always isotropic to order $d^{-3}$. The anisotropic terms are ruled out after matching boundary condition at the interface.

In this comment we show that the assumption that $\phi^m(k)$ is always a regular function at small *k* in expansion (1) is invalid in general, and as a result, anisotropic (dipolar) interaction between colloids can exist up to order $d^{-3}$, in contrast to the claim by Domínguez *et. al.*

To illustrate we consider the electrostatic problem shown in the figure below, which is a system with cylindrical symmetry and the figure shows a side view.

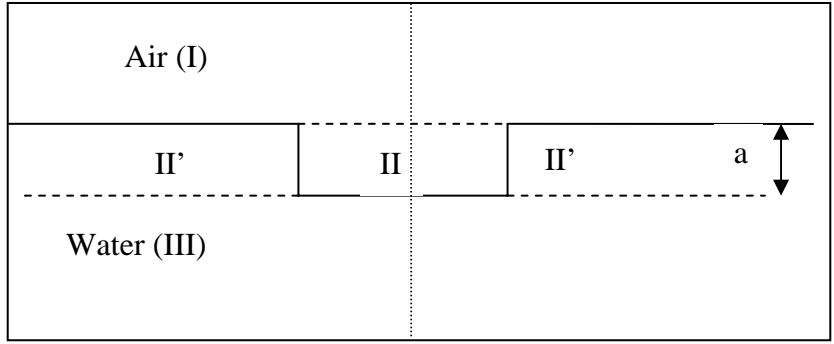

The electrostatic potential is determined by the Poisson equation $\nabla^2 \phi = 0$ at regions (I) and (II), and the Poisson-Boltzmann equation at region (II') and (III). To simplify the problem we shall replace region II by a "liquid" with a dielectric constant differ from water, i.e. Region II + II' can be described by the Poisson-Boltzmann equation,

$$(-\nabla^2 + \kappa^2)\phi = 0, \qquad (3)$$

except that $\partial_r \phi$ is discontinuous across the boundary between II and II'. To make the problem non-trivial, we may place a surface charge density $\sigma(\vec{r})$ on the surface between regions I and II. We shall assume further that the thickness $a$ of region II satisfies $a \gg \kappa^{-1}$, so that we can neglect region III when we solve the Poisson-Boltzmann equation.

In this case the general solution in Region II + II' is simple and is given by

$$\phi_{II}(r,z,\vartheta) = \sum_m \int dk\, \phi_{II}^m(k) R_m(kr) e^{Kz} e^{im\vartheta}, \qquad (4)$$

where $K = \sqrt{k^2 + \kappa^2}$ [1] and the only difference between the present problem and the totally flat surface is that the Bessel function $J_m(kr)$ is replaced by the function $R_m(kr)$ which is the solution of the corresponding scattering problem. In the above simple situation,

$$R_m(kr) \sim J_m(kr) \qquad \text{(Region II)} \qquad (5)$$
$$R_m(kr) \sim a_{km} J_m(kr) + b_{km} N_m(kr) \qquad \text{(Region II')}$$

where $N_m(kr)$ is the Neumann function. Similarly region I is described by

$$\phi_I(r,z,\vartheta) = \sum_m \int dk\, \phi_I^m(k) J_m(kr) e^{-kz} e^{im\vartheta}. \qquad (6)$$

$\phi_{I(II)}^{m}(k)$'s are determined by matching the solution at the boundary between I and (II + II'). We note that we can always write

$$R_m(kr) = \int dk' A(k,k') J_m(k'r) \tag{7a}$$

because of completeness, where

$$A(k,k') = k' \int r\, dr\, J_m(kr) R_m(kr). \tag{7b}$$

Therefore we may also write,

$$\phi_{II}(r,0,\vartheta) = \sum_m \int dk\, \overline{\phi}_{II}^{m}(k) J_m(kr) e^{im\vartheta}, \tag{8a}$$

where

$$\overline{\phi}_{II}^{m}(k) = \int dk' A(k',k) \phi_{II}^{m}(k') = \phi_{I}^{m}(k). \tag{8b}$$

The last equality is a result of the boundary condition $\phi_I(r,z=0,\vartheta) = \phi_{II}(r,z=0,\vartheta)$ at the boundary.

Now we want to show that in general $\phi_I^m(k)$ and/or $\phi_{II}^m(k)$ are singular at small $k$ or $A(k,k')$ is in general singular at small $k$, $k'$.

To show this we examine the behavior of

$$\phi_{I(II)}(r,0,\vartheta) = \sum_m \int dk\, \phi_I^m(k) J_m(kr) e^{im\vartheta} = \sum_m \int dk\, \phi_{II}^m(k) R_m(kr) e^{im\vartheta} \tag{9}$$

at large $r$.

Assuming $\phi_I^m(k)$ is regular at small $k$, we may follow ref.[1] and extract the large $r$ behavior from the coefficients $\phi_{J,j}^m = \sum_m \int dk\, k^j J_m(kr)$ which behave as

$$\phi_{I,j}^m = \phi_{J,j}^m \sim \begin{cases} r^{-j-1} e^{im\vartheta} & |m| > j \ or \ j-|m| \ even \\ 0 & otherwise \end{cases} \tag{10}$$

However, we may also perform the same analysis for $\phi_{II}^m(k)$ by looking at the long distance behavior of $R_m(kr) \sim a_{km} J_m(kr) + b_{km} N_m(kr)$ which is determined by the coefficients $\phi_{J,j}^m$ and

$$\phi_{N,j}^m = \sum_m \int dk k^j N_m(kr). \tag{11}$$

In particular, we find that [3]

$$\lim_{h \to 0} \int_0^\infty dk k^j N_m(kr) e^{-kh} = \frac{2^j}{r^{j+1}} \sin\left(\frac{(j-|m|)\pi}{2}\right) \Gamma\left(\frac{|m|+j+1}{2}\right) \Gamma\left(\frac{j+1-|m|}{2}\right)^{-1}$$

$$\sim \begin{cases} r^{-j-1} & j+1 > |m|, \ j-|m| \ odd \\ 0 & otherwise \end{cases} \tag{12}$$

and, together with $\phi_{J,j}^m$ we obtain

$$\phi_{II,j}^m \sim a_{km} \phi_{J,j}^m + b_{km} \phi_{N,j}^m \sim r^{-j-1} e^{im\vartheta} \tag{13}$$

for all possible values of $m$ and $j$. The boundary matching condition Eq.(9) requires that

$$\phi_{I,j}^m \equiv \phi_{II,j}^m \tag{14}$$

for all $m$ and $j$, which is impossible in general if we assume that $a(b)_{km}$ and $\phi_{I(II)}^m(k)$ are are regular at small $k$. $a(b)_{km}$ are determined by solving the Possion-Boltzmann and are generally regular. Therefore $\phi_I^m(k)$ and/or $\phi_{II}^m(k)$ must become singular at small $k$ if Eq.(14) is satisfied. In the above simple example, the singularity can be traced back to the singularity in $A(k,k')$ coming from the discontinuity $\partial_r R_m(kr)$ across the boundary between Region II and II'. The unavoidable singularity in the above example invalidates the argument in Ref.[1] which depends on the regularity of the coefficients.

It is not difficult to show that the boundary matching of $\frac{\partial \phi}{\partial z}$ across the boundary also leads to the same conclusion, and in general anisotropic $d^{-3}$ interaction between colloids on interfaces are permitted to exist when we take into account the non-analytic behavior of $\phi_I^m(k)$ and/or $\phi_{II}^m(k)$, in contract to what is claimed in Ref.[1].

Summarizing, when solving the electrostatic problem in the presence of local disturbances, it is very dangerous to assume that the surface potential at long distance can be expanded in a form $\phi(r, z=0, \vartheta) = \sum_m \int dk \phi^m(k) J_m(kr) e^{im\vartheta}$ with regular coefficients $\phi^m(k)$'s. We have demonstrated this explicitly using one simple example in this comment. In fact it can be seen by simply searching Ref.[3] that there are more than one examples of regular functions $f(r,\vartheta)$ which, when expanded in terms of Bessel functions $J_m(kr)$, i.e. $f(r,\vartheta)=\sum_m\int dk \bar{f}^m(k)J_m(kr)e^{im\vartheta}$, admitted singular coefficients $\bar{f}^m(k)$'s. Therefore we believe that the claim in Ref.[1] that the electrostatic interaction between charged colloids trapped at an interface formed by a dielectric and a screening phase is always isotropic to order $d^{-3}$ is incorrect, and the problem should be handled more carefully.